\documentclass{PoS}
\title{Transversity Parton Distribution}

\ShortTitle{Transversity Parton Distribution}

\author{\speaker{Alexei Prokudin}\\%\thanks{A footnote may follow.}\\
        Jefferson Lab, 12000 Jefferson Avenue, Newport News, Virginia
  23606, USA\\
        E-mail: \email{prokudin@jlab.org}}

\abstract{Transversity distribution is one of the three fundamental parton distributions that completely describe
polarized spin 1/2 nucleon. Its chiral odd nature prevented for many years its experimental exploration, however presently
we have obtained great deal of information about this distribution. This includes experimental data from Semi Inclusive Deep Inelastic Scattering, knowledge of scale dependence and phenomenological extractions. I will discuss main features of this distribution and indicate the future improvements of our knowledge.}

\FullConference{The 7th International Workshop on Chiral Dynamics,\\
		August 6 -10, 2012\\
		Jefferson Lab, Newport News, Virginia, USA}

\maketitle

\begin{document}

\section{Introduction}

Distribution of partons in a polarized spin-1/2 hadron can be completely described by three
collinear Parton Distribution Functions (PDFs): unpolarised parton distribution $f_1$, helicity distribution $g_1$, and
transversity distribution $h_1$. These standard collinear PDFs are defined
through collinear factorization theorems and obey DGLAP
evolution equations~\cite{Altarelli:1977zs,Dokshitzer:1977sg,Lipatov:1974qm,Barone:1997fh,Vogelsang:1997ak,Hayashigaki:1997dn}. 

The transversity distribution \cite{Ralston:1979ys} describes transversely polarized quarks in a  transversely polarized nucleon. Formally such a
distribution can be expressed by
\begin{equation}
  h_1^q(x) = \int \frac{d\xi^-}{2\pi} e^{-ix P^+ \xi^-}\langle P, S_{T}| {\bar{\psi}}_q(0^+,\xi^-,0_\perp) \frac{\gamma^+ \gamma^j \gamma_5}{2}\psi_q(0^+,0^-,0_\perp)|P,S_{T}\rangle
\end{equation}
\noindent 
The corresponding charge is called ``tensor charge''
\begin{equation}
 \delta q = \int_0^1 d x (h_1^q(x) - h_1^{\bar q}(x))
\end{equation}
\noindent
Transversity obeys so-called Soffer bound \cite{Soffer:1995ww}
\begin{equation}
 |h_1(x,Q^2)| \le \frac{1}{2}\left(f_1(x,Q^2) + g_1(x,Q^2) \right)\,
\end{equation}
This bound  was shown to be preserved at LO accuracy in
Ref.~\cite{Barone:1997fh} and at NLO accuracy in
Ref.~\cite{Vogelsang:1997ak}.  
 
Transversity is the least known of the three collinear distributions and the reason is that
it cannot be measured in Deep Inelastic Scattering (DIS) due to its chiral odd nature. It should couple to another chiral odd 
quantity (chiral odd fragmentation or chiral odd distribution function, for example transversity itself). The best channel
to measure transversity remains polarized Drell-Yan (preferable proton anti proton) process in which one could measure the product
of transversity distributions directly, see Ref.~\cite{Barone:2005pu}. QCD evolution of collinear transversity distribution is well known, see Refs.~\cite{Barone:1997fh,Vogelsang:1997ak,Hayashigaki:1997dn}. It does not couple to gluons and thus exhibits non-singlet $Q^2$ evolution. Gluon transversity distribution does not exist either. This leads to the fact that transversity is suppressed at low-x makes it
a valid object to study in high-x region by Jefferson Lab 12 \cite{Dudek:2012vr}.

Currently the knowledge on transversity comes from Semi Inclusive Deep Inelastic Scattering (SIDIS) experimentally observed at HERMES \cite{Airapetian:2004tw,
Airapetian:2010ds}, COMPASS \cite{Ageev:2006da,Martin:2013eja} and JLab 6 \cite{Qian:2011py} in single spin asymmetries where transversity couples to so-called Collins fragmentation function~\cite{Collins:1992kk}. Information on the convolution of two chiral-odd fragmentation functions 
 is obtained from $e^+e^- \to h_1 \, h_2 \, X$ processes ~\cite{Abe:2005zx,Seidl:2008xc,Seidl:2012er}.    One usually measures low transverse momentum final hadron and thus one applies Transverse Momentum Dependent factorization. The transversity in this case depends also on intrinsic transverse 
motion of quarks $\bf k_\perp$ and one speaks of Transverse Momentum Dependent (TMD) transversity. One can also study transversity coupled to so-called di-hadron fragmentation function~\cite{Collins:1993kq, Jaffe:1997hf,
Radici:2001na}.

The $u$ and $d$ quark transversity distributions, together with the Collins 
fragmentation functions, have been extracted for the first time in 
Refs.~\cite{Anselmino:2007fs, Anselmino:2008jk}, from a combined analysis 
of SIDIS and $e^+e^-$ data. The most recent extraction is presented in Ref.~\cite{Anselmino:2013vqa}.

Di-hadron method was implemented in the analysis of Ref.~\cite{Bacchetta:2012ty} and the results
on the extraction of transversity from Refs.~\cite{Anselmino:2007fs, Anselmino:2008jk,Anselmino:2013vqa}
and Ref.~\cite{Bacchetta:2012ty} agree with each other quite well. 

QCD evolution of Transverse Momentum Dependent transversity was recently obtained in Ref.~\cite{Bacchetta:2013pqa}.

\section{Phenomenology}

\begin{figure}[th]
\begin{center}
\vskip -1.cm
\includegraphics[width=0.5\textwidth, angle=-90]
{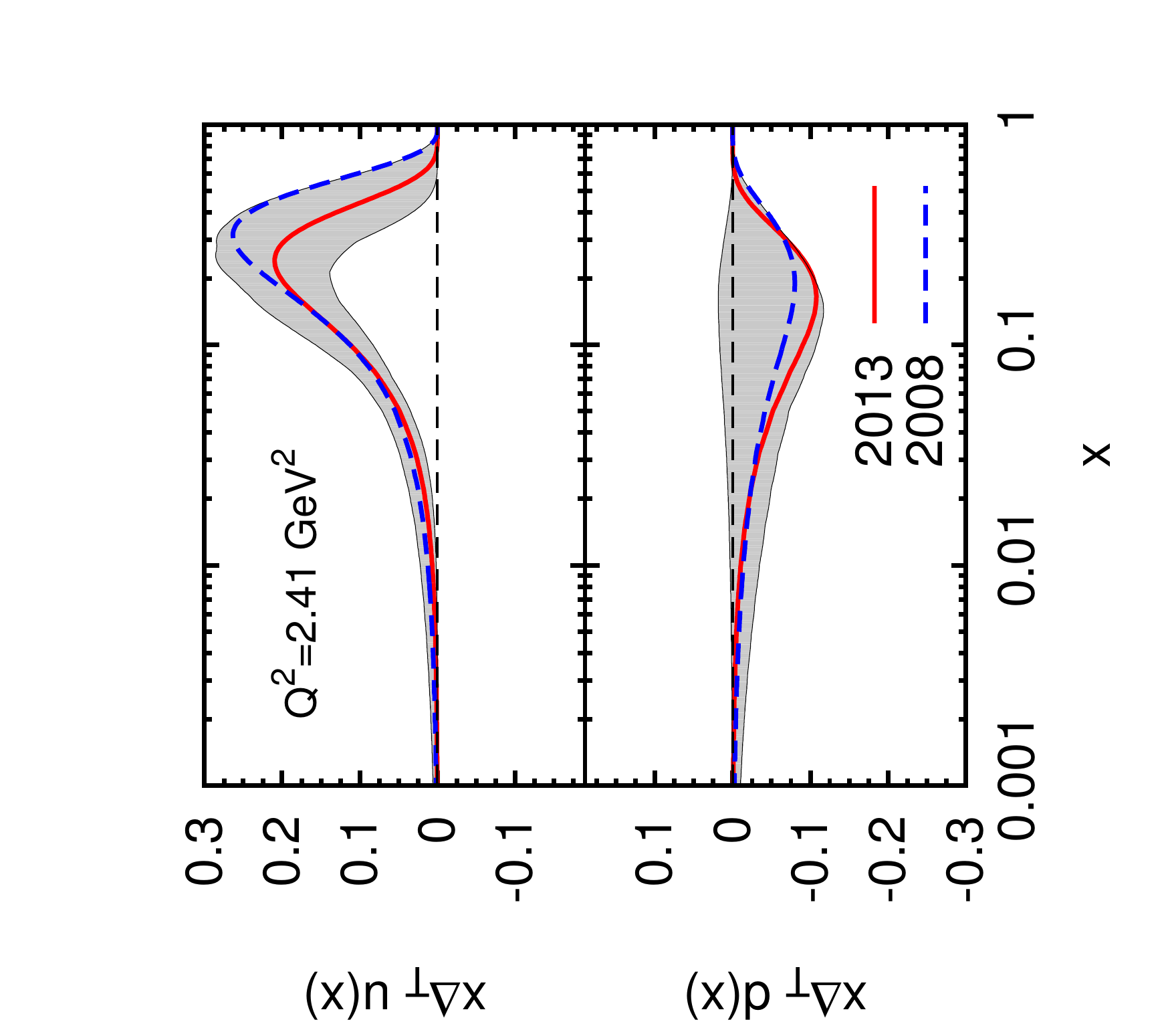}
\vskip -0.5cm
\caption{\label{fig:newh1-collins-A12}
The transversity 
distribution functions $x \, h_{1}^{q}(x) = x \, \Delta_T q(x)$ for $q = u,d$,
with their uncertainty bands (shaded areas), obtained from the best fit 
of SIDIS data   and $e^+e^-$ data in Ref.~\cite{Anselmino:2013vqa}.
The dashed blue lines show the same quantities as obtained in 
Ref.~\cite{Anselmino:2008jk}. 
}
\end{center}
\end{figure}

The result on extraction of the transversity is presented in Fig.~\ref{fig:newh1-collins-A12}.\footnote{The plot is from Ref.~\cite{Anselmino:2013vqa}} One can see that $u$ quark transversity is positive and $d$ quark transversity is negative.
This results is coming from global analysis of SIDIS HERMES \cite{Airapetian:2004tw,
Airapetian:2010ds}, COMPASS \cite{Ageev:2006da,Martin:2013eja} and $e^+e^-$ BELLE \cite{Abe:2005zx,Seidl:2008xc,Seidl:2012er} data.

Experimentally so-called Collins asymmetry in SIDIS with unpolarised beams ($U$) and transversely polarized target ($T$) is measured and it is proportional to convolution of transversity and Collins fragmentation functions
\begin{equation}
 A_{UT}^{\sin(\phi_h+\phi_S)} \propto \sum_q h_1^q\otimes H_{1q}^\perp
\end{equation}
 here $\phi_h$, and $\phi_S$ are azimuthal angles of produced pion and polarization vector, experimentally observed
modulation is proportional to $\sin(\phi_h+\phi_S)$ and sign $\otimes$ denotes usual TMD convolution \cite{Collins:2011zzd}.

One can see that knowledge of Collins fragmentation function ($H_{1q}^\perp$) is needed in order to extract transversity, fortunately 
in $e^+e^-$ process one observes an asymmetry which is related to convolution of two Collins functions $\sum_q H_{1q}^\perp \otimes H_{1\bar q}^\perp$. This allows us to have perform global analysis ~\cite{Anselmino:2007fs, Anselmino:2008jk,Anselmino:2013vqa} of SIDIS and $e^+e^-$ data.

%%%%%%%%%%%%%%%%%%%%%%%%%%%%%%%%%%%%%%%%%%%%%%%%%%%%%%%%%%%%%%%%%%%%%%%%%%%
% Transversity e Collins functions - A_12 Standard
%%%%%%%%%%%%%%%%%%%%%%%%%%%%%%%%%%%%%%%%%%%%%%%%%%%%%%%%%%%%%%%%%%%%%%%%%%%%%%%
\begin{figure}[th]
\begin{center}
\vskip -0.5cm
\includegraphics[width=0.9\textwidth, angle=0]{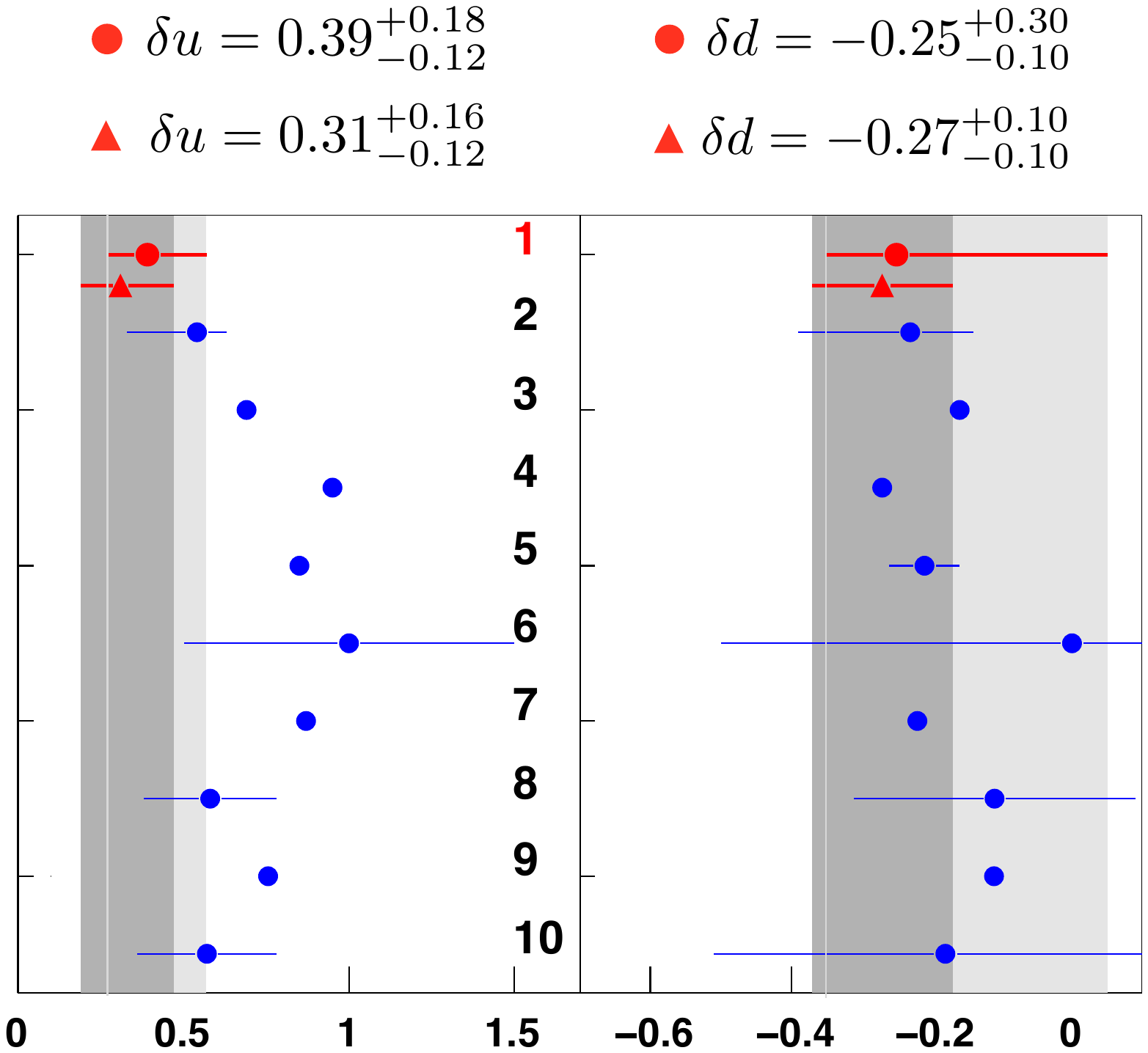}
\vskip -0.5cm
\vskip -36pt
\caption{\label{fig:tensorcharge}
The tensor charge  
for $u$ (left) and $d$ (right) quarks, computed using the transversity 
distributions obtained in Ref.~\cite{Anselmino:2013vqa}. The gray areas 
correspond to the statistical uncertainty bands of the extraction. The 
results are compared with those given in Ref.~\cite{Anselmino:2008jk} 
(number 2)  of the model calculations of 
Refs.~\cite{Cloet:2007em,Wakamatsu:2007nc,Gockeler:2005cj,He:1994gz,
Pasquini:2006iv,Gamberg:2001qc,Hecht:2001ry}
(number 3-9) and with the results extraction from Ref.~\cite{Bacchetta:2012ty} (number 10).}
\end{center}
\end{figure}

%%%%%%%%%%%%%%%%%%%%%%%%%%%%%%%%%%%%%%%%%%%%%%%%%%%%%%%%%%%%%%%%%%%%%%%%%%
\noindent
We also present the results on the tensor charge at $Q^2=$ 0.8 GeV$^2$ in Fig.~\ref{fig:tensorcharge}.\footnote{The plot is from Ref.~\cite{Anselmino:2013vqa}}

%%%%%%%%%%%%%%%%%%%%%%%%%%%%%%%%%%%%%%%%%%%%%%%%%%%%%%%%%%%%%%%%%%%%%%%%%%%%%%%%
% tensor charge 
%%%%%%%%%%%%%%%%%%%%%%%%%%%%%%%%%%%%%%%%%%%%%%%%%%%%%%%%%%%%%%%%%%%%%%%%%%%%%%%

Future measurements at Jefferson Lab 12 are going to be very important for the extraction of the transversity and the
tensor charge. We estimate that corresponding improvement of the statistical error of the extraction will be a factor
of $5$ approximately~\cite{Dudek:2012vr}. This will mean that from almost 50\% uncertainty we will be able to extract
tensor charge with 10\% uncertainty and compare it better to model predictions.

Tensor charge is important for some dynamical effects of new heavy Beyond Standard Model degrees of freedom \cite{Bhattacharya:2011qm}. In order to constrain
possible parameters of those models one needs precise knowledge of the tensor charge.

Future Electron Ion Collider \cite{Boer:2011fh} will also allow us to study carefully $Q^2$ evolution of transversity and explore
low-x region.

\section{Conclusions}

Interested reader is referred to several reviews that describe the transversity in greater detail, see Refs~\cite{Barone:2001sp,Barone:2010zz,D'Alesio:2007jt}. I have not discussed all possible ways to access transversity,
for instance $\Lambda$ electroproduction in SIDIS. In this case one needs to know the chiral odd fragmentation function
of $\Lambda$ production and it can be accessed via $e^+e^- \rightarrow \bar\Lambda\Lambda X$. One could also study transversity
in proton proton scattering by utilizing $ p p^\uparrow \rightarrow \pi jet X$ \cite{D'Alesio:2010am}.

In future we will have data from BABAR 
Collaboration, which have performed an independent new analysis of 
$e^+ e^- \to h_1 \, h_2 \, X$ data~\cite{Garzia:2012za}. Jefferson Lab 12 will provide precision data in high-x region \cite{Dudek:2012vr} and thus complement results obtained in SIDIS at HERMES \cite{Airapetian:2004tw,
Airapetian:2010ds}, COMPASS \cite{Ageev:2006da,Martin:2013eja} and JLab 6 \cite{Qian:2011py}.

Generally proton proton scattering can be described by so-called twist-3 factorization in which one studies multi-parton correlations
of partons and corresponding Efremov-Teryaev-Qiu-Sterman functions \cite{Efremov:1981sh,Efremov:1984ip,Qiu:1991pp,Qiu:1998ia,Koike:2009ge,Kang:2010zzb}. These functions are related to TMD functions and more globally twist-3 formalism and TMD formalism are closely related to each other, and have been shown to be equivalent in the overlap region where both can apply \cite{Ji:2006ub,Koike:2007dg,Bacchetta:2008xw}.

Once a comprehensive global analysis of the data from SIDIS (TMD) and proton proton scattering (twist-3) is done (see preliminary
results in Refs.~\cite{Kang:2012xf,Gamberg:2013kla}) we are going to obtain a complete description of asymmetries and corresponding parton distributions
including transversity parton distribution.

%\section{Acknowledgements}
{\bf Acknowledgements} The author would like to thank his colleagues 
Mauro Anselmino, Elena Boglione, Umberto D'Alesio, Stefano Melis, Francesco Murgia, Leonard Gamberg, and Zhong-Bo Kang.
The main results presented in this article are obtained in collaboration with them. 
  
%\bibliographystyle{is-abbrv}
%\bibliographystyle{is-unsrt}
%\bibliographystyle{amsplain}
%\bibliographystyle{JHEP-2.bst}
%\bibliography{mybiblio}

%\begin{thebibliography}{99}
%\bibitem{...} 
%....

%\end{thebibliography}
 \providecommand{\href}[2]{#2}\begingroup\raggedright\endgroup

\end{document}